\documentclass[12pt,a4paper]{article}
\usepackage[british]{babel}
\usepackage{epsfig}
\begin{document}
\date{}

\title{
{\vspace{-20mm} \normalsize
\hfill \parbox[t]{50mm}{DESY 04-046  \\
                        MS-TP-04-3}} \\[20mm]
 Quark mass dependence of pseudoscalar masses and \\
 decay constants on a lattice}

\author{qq+q Collaboration                                  \\[0.5em]
        F. Farchioni                                        \\[0.5em]
        Westf\"alische Wilhelms-Universit\"at M\"unster,    \\
        Institut f\"ur Theoretische Physik,                 \\
        Wilhelm-Klemm-Strasse 9, D-48149 M\"unster, Germany \\[1em]
        I. Montvay, E. Scholz                               \\[0.5em]
        Deutsches Elektronen-Synchrotron DESY               \\
        Notkestr.\,85, D-22603 Hamburg, Germany}

\newcommand{\be}{\begin{equation}}                                              
\newcommand{\ee}{\end{equation}}                                                
\newcommand{\half}{\frac{1}{2}}                                                 
\newcommand{\rar}{\rightarrow}                                                  
\newcommand{\lar}{\leftarrow}
\newcommand{\LCB}{\raisebox{-0.3ex}{\mbox{\LARGE$\left\{\right.$}}}
\newcommand{\RCB}{\raisebox{-0.3ex}{\mbox{\LARGE$\left.\right\}$}}}
\newcommand{\U}{\mathrm{U}}
\newcommand{\SU}{\mathrm{SU}}
\newcommand{\bteq}[1]{\boldmath$#1$\unboldmath}


\maketitle

\abstract{
 Our previous calculations of the sea- and valence-quark mass dependence
 of the pseudoscalar meson masses and decay constants is repeated on a
 $16^3 \cdot 32$ lattice which allows for a better determination of
 the quantities in question.
 The conclusions are similar as before on the $16^4$ lattice
 \protect{\cite{SEA}}.
 The two light dynamical quark flavours we simulate have masses in the
 range $\frac{1}{4}m_s < m_{u,d} < \frac{2}{3}m_s$. 
 The sea quark mass dependence of $f_\pi$ and $m_\pi^2/m_q$ is well
 described by the next-to-leading order (NLO) Chiral Perturbation Theory
 (ChPT) formulas and clearly shows the presence of chiral logarithms.
 The valence quark mass dependence requires the presence of NNLO
 contributions in Partially Quenched ChPT (PQChPT) -- in addition to
 the NLO terms.
 The ${\cal O}(a)$ lattice artifacts in these quantities turn out to be
 small.}

\newpage
\section{Introduction}\label{sec1}

 In Quantum Chromodynamics (QCD) -- the theory of strong interactions --
 there are two very light quarks and one moderately light quark ($u,d$
 and $s$, respectively).
 The strong interaction dynamics at low energies can be formulated by
 an {\em effective chiral Lagrangian} which incorporates the symmetry
 constraints following from the spontaneously broken chiral symmetry of
 the light quarks.
 In this low energy effective theory the interactions are described by
 a simultaneous expansion in powers of momenta and light quark masses
 \cite{WEINBERG,CHPT}.
 The coefficients of the interaction terms in the effective chiral
 Lagrangian -- the {\em Gasser-Leutwyler constants} -- are free
 parameters which can be constrained by experimental data and also
 calculated from the underlying basic QCD Lagrangian in the framework
 of the non-perturbative lattice regularization.

 In numerical lattice QCD simulations the quark masses are free
 parameters.
 Changing these parameters gives an excellent opportunity to precisely
 determine the Gasser-Leutwyler constants.
 In fact, ChPT based on the chiral Lagrangian can be extended by
 changing the {\em valence quark masses} in quark propagators
 independently from the {\em sea quark masses} in virtual quark loops.
 This leads to partially quenched chiral perturbation theory (PQChPT)
 \cite{BERNARD-GOLT}.

 The aim of numerical simulations in QCD is to reach the regime of light
 quark masses where NLO chiral perturbation theory gives a good
 approximation.
 In previous papers \cite{NF2TEST,VALENCE,SEA} our collaboration started
 a series of simulations with two equal-mass light quarks ($qq$) with
 the goal of extracting the values of the Gasser-Leutwyler constants
 conventionally denoted by $L_k,\; (k=1,2,\ldots)$.
 Later on it will be possible to extend these calculations by also
 including the $s$-quark ($qq$+$q$).

 In our previous paper \cite{SEA} we started some larger scale
 simulations on a $16^4$ lattice at the gauge coupling $\beta=5.1$ which
 corresponds to a lattice spacing of $a \simeq 0.2\,{\rm fm}$.
 Since it became clear that interesting results can be obtained already
 at this relatively rough discretization scale, we decided to repeat
 and extend these simulations on a $16^3 \cdot 32$ lattice which is
 better suited for extracting quantities like the pseudoscalar
 (``pion'') mass ($m_\pi$) and decay constant ($f_\pi$).
 Our work profited from the valuable experience of previous simulations
 by other collaborations \cite{ALPHA:CHPT,UKQCD:CHPT,STAGGERED:CHPT}.

 Since the present work is on the same topics as ref.~\cite{SEA}, we
 shall often only refer to it without repeating its full content.
 In general, we use the conventions and notations of
 \cite{SEA,VALENCE,NF2TEST}.
 Nevertheless, we also try to make the present paper easily
 understandable for the reader and therefore repeat the main definitions
 and relations.
 In the next section we deal with the sea quark mass dependence of
 $f_\pi$ and $m_\pi$.
 In section \ref{sec3} the valence quark mass dependence is considered
 and the question of the magnitude of leading lattice artifacts is
 investigated.
 Section \ref{sec4} is a short summary of our experience with the
 Monte Carlo updating algorithm.
 The last section contains the summary and discussion.

\section{Sea quark mass dependence}\label{sec2}

 We performed Monte Carlo simulations with $N_s=2$ degenerate sea quarks
 on a $16^3 \cdot32$ lattice at gauge coupling $\beta=5.1$ and four
 values of the hopping parameter $\kappa$:
 $\kappa_0=0.176$, $\kappa_1=0.1765$, $\kappa_2=0.1768$ and
 $\kappa_3=0.177$.
 Three of these points have also been simulated previously on the
 $16^4$ lattice in ref.~\cite{SEA}.
 The point at $\kappa_1=0.1768$ is new.
 We collected 950-1000 gauge configurations per point which are
 typically separated by 10 update cycles consisting out of boson
 field and gauge field updates and noisy correction steps.
 (Some observations about the algorithm will be summarized in section
 \ref{sec4}.)

 A collection of the values of some basic quantities in these simulation
 points is given in table \ref{tab1}: the Sommer scale-parameter in
 lattice units $r_0/a$, the pion mass in lattice units $am_\pi$,
 the quark mass parameter $M_r=(r_0 m_\pi)^2$, the bare PCAC quark mass
 $Z_q am_q$ including the multiplicative renormalization factor
 $Z_q=Z_P/Z_A$, the ratio of the PCAC quark masses $\sigma_i$ with
 respect to the {\em reference sea quark mass} at $\kappa=\kappa_0$ and
 the pion decay constant in lattice units $af_\pi$ divided by the
 renormalization factor $Z_A$.
 (The normalization of the pion decay constant is such that the physical
 value is $f_\pi \simeq 93\, {\rm MeV}$.)
\begin{table}[ht]
\begin{center}
\parbox{12cm}{\caption{\label{tab1}\em
 The values of some basic quantities in our simulation points.
 Statistical errors in last digits are given in parentheses.}}
\end{center}
\begin{center}
\begin{tabular}[ht]{|c||c|c|c|c|}
\hline
$\kappa$                 & $\kappa_0$  & $\kappa_1$  & $\kappa_2$  & $\kappa_3$
\\ \hline\hline
$r_0/a$                  & 2.229(63)   & 2.212(44)   & 2.621(46)   & 2.528(51)
\\ \hline
$am_\pi$                 & 0.6542(10)  & 0.5793(17)  & 0.3919(46)  & 0.3657(24)
\\ \hline
$M_r=(r_0 m_\pi)^2$      & 2.13(12)    & 1.642(72)   & 1.055(36)   & 0.855(34)
\\ \hline
$Z_q am_q$               & 0.07092(27) & 0.05571(30) & 0.02566(27) & 0.02208(21)
\\ \hline
$\sigma_i=m_{qi}/m_{q0}$ & 1.0         & 0.7856(56)  & 0.3618(44)  & 0.3113(31)
\\ \hline
$Z_A^{-1} af_\pi$        & 0.2819(15)  & 0.2590(14)  & 0.2008(17)  & 0.1936(16)
\\ \hline
\end{tabular}
\end{center}
\end{table}

 Comparing table \ref{tab1} to the corresponding one (table 3) in
 \cite{SEA} one can see that these quantities extracted on the
 $16^3 \cdot 32$ lattice differ considerably from those extracted on
 the $16^4$ lattice.
 The change of $r_0/a$ is about 2-5\%.
 The difference in $am_\pi$ increases from 3\% at $\kappa_0$ to about
 16\% at $\kappa_3$ whereas $Z_q am_q$ differs at $\kappa_0$ by 5\% and
 at $\kappa_3$ already by about 28\%.
 However, as we shall see later on, considering ratios of the pion
 mass-square and of the pion decay constant as a function of the ratios
 of PCAC quark masses (denoted by $\sigma$ for sea quark masses and
 $\xi$ for valence quark masses) it turns out that almost all changes
 between the $16^4$ and $16^3 \cdot 32$ lattices cancel.

 In table \ref{tab1} the bare quark mass obtained from the PCAC relation
 is shown: $m_q \equiv m_q^{PCAC}$.
 (For details of its numerical determination see section 3.1.1 in
 ref.~\cite{NF2TEST}.)
 Another possibility to define the quark mass is to take
 $am_{ren}\equiv \mu_{ren} \equiv Z_m(\mu_0-\mu_{cr})$ where
 $\mu_0 = 1/(2\kappa)-4$ is the
 bare quark mass in the Wilson-fermion action, $\mu_{cr}$ is its
 critical value corresponding to zero quark mass and $Z_R$ is an
 appropriate multiplicative renormalization factor.
 The values of $\mu_0$ corresponding to $\kappa_0,...,\kappa_3$ are
 $\mu_{0(0)}=-1.1590909..,\; \mu_{0(1)}=-1.1671388..,\;
  \mu_{0(2)}=-1.1719457..,\; \mu_{0(3)}=-1.1751412..$, respectively.
 Comparing the values of $Z_q am_q$ or $\sigma_i$ in table \ref{tab1}
 to the values $\mu_{0(i)}$ one can see that the relation between them
 is highly non-linear.
 This implies the same also for the relation between the (ratios of)
 $m_q^{PCAC}$ and $m_{ren}$.
 The non-linear terms in this relation are lattice artifacts which
 have to vanish in the continuum limit but they are large at our
 lattice spacings.

 A consequence of the strongly non-linear relation between $\sigma$ and
 $\mu_0$ is that the determination of $\mu_{cr}$ (or $\kappa_{cr}$)
 has large uncertainty.
 In fact, with our four points only we could not find a convincing
 extrapolation of $\sigma$ to zero.
 A crude quadratic extrapolation gives $\mu_{cr}=-1.180(4)$ or
 $\kappa_{cr}=0.1773(2)$.
 The uncertainty in the critical point implies an uncertainty in the
 extrapolation of physical quantities, too, which is necessary in
 a quark mass independent renormalization scheme.
 In case of the lattice spacing, which can be obtained from the
 extrapolation of $r_0/a$ to the critical point, table \ref{tab1} shows
 that the values of $r_0/a$ increase between $\kappa_0$ and $\kappa_2$
 but between $\kappa_2$ and $\kappa_3$ they are within errors constant.
 Therefore we take this constant value as the extrapolated one:
 $[r_0/a]_{cr} = 2.57(5)$.
 This gives, with $r_0 \equiv 0.5\,{\rm fm}$, for the quark mass
 independent lattice spacing $a=0.195(4)\,{\rm fm}$.

 In the ChPT formulas the quark mass can be represented by the
 dimensionless quantity
\be\label{eq2-1:01}
\chi \equiv \frac{2B_0 m_q}{f_0^2}
\ee
 where $B_0$ is a conventional parameter with dimension mass and
 $f_0$ is the value of the pion decay constant at zero quark mass.
 (Its normalization here is such that the physical value is
 $f_0 \simeq 93\, {\rm MeV}$.)
 In what follows we shall identify the quark mass $m_q$ in $\chi$ with
 the PCAC quark mass $m_q^{PCAC}$.
 According to the previous discussion this is a non-trivial choice
 because the lattice artifacts in (ratios of) the quark mass are
 rather different for $am_q^{PCAC}$ then, for instance, for $am_{ren}$.

 The sea quark mass dependence of the ratio of the pion decay constant
 in NLO of ChPT is:
\be\label{eq2-2:02}
Rf_{SS} \equiv \frac{f_{SS}}{f_{RR}} =
1 + 4(\sigma-1)\chi_R (N_s L_{R4}+L_{R5}) -
\frac{N_s\chi_R}{32\pi^2}\sigma\log\sigma + {\cal O}(\chi_R^2) \ .
\ee
 Here $f_{SS}$ is the pion decay constant of a pion consisting of
 two sea quarks with mass $\chi_S$ and $f_{RR}$ is its value at some
 {\em reference quark mass} $\chi_R$.
 $N_s$ is the number of mass-degenerate sea quarks (actually $N_s=2$),
 $L_{Rk}\; (k=4,5,\ldots)$ are Gasser-Leutwyler constants at the
 scale $\mu=f_0\sqrt{\chi_R}$ and the ratio of sea quark masses to the
 reference quark mass is
\be\label{eq2-3:03}
\sigma   \equiv \frac{\chi_S}{\chi_R} \ .
\ee
 The analogous formula for the pion mass squares is:
\begin{eqnarray}
Rn_{SS} \equiv \frac{m_{SS}^2}{\sigma m_{RR}^2} =
1 &+& 8(\sigma-1)\chi_R(2N_s L_{R6}+2L_{R8}-N_s L_{R4}-L_{R5})
\nonumber
\\[0.5em]\label{eq2-4:04}
&+& \frac{\chi_R}{16\pi^2 N_s}\sigma\log\sigma + {\cal O}(\chi_R^2) \ .
\end{eqnarray}
 Note that instead of the scale dependent combinations (at $N_s=2$)
\be\label{eq2-5:05}
L_{R45} \equiv 2L_{R4}+L_{R5} \ , \hspace{2em}
L_{R6845} \equiv 4L_{R6}+2L_{R8}-2L_{R4}-L_{R5}
\ee
 one can also use the {\em universal low energy scales} $\Lambda_{3,4}$
 defined by \cite{LEUTWYLER}
\begin{eqnarray}
\Lambda_3 &=& 4\pi f_0\exp(-\alpha_{6845}) \ , \hspace{3em}
\alpha_{6845} = 128\pi^2 L_{R6845}-\half\log\frac{\chi_R}{16\pi^2}
\nonumber
\\[1.0em]\label{eq2-6:06}
\Lambda_4 &=& 4\pi f_0\exp(\alpha_{45}/4) \ , \hspace{4.4em}
\alpha_{45} = 128\pi^2 L_{R45}+2\log\frac{\chi_R}{16\pi^2} \ .
\end{eqnarray}

 The free parameters in $Rf_{SS}$ and $Rn_{SS}$ are $\chi_R$,
 $\chi_R L_{R45}$ and $\chi_R L_{R6845}$.
 With the small number of points we have the linear fit with these
 parameters gives a good chi-square but relatively large errors:
 $\chi^2=0.8$ and
\be\label{eq2-7:07}
\chi_R = 30.8(9.4) \ ,                \hspace{2em}
\chi_R L_{R45} = 0.1398(86) \ ,       \hspace{2em}
\chi_R L_{R6845} = -0.0078(22) \ .
\ee
 This corresponds to
\begin{eqnarray}
&& L_{R45} = 4.5(1.1) \cdot 10^{-3} \ ,   \hspace{3em}
\frac{\Lambda_4}{f_0} = 23.3(8.2)\ ,
\nonumber
\\[0.5em]\label{eq2-8:08}
&& L_{R6845} = -2.54(21) \cdot 10^{-4} \ , \hspace{3em}
\frac{\Lambda_3}{f_0} = 7.64(14) \ .
\end{eqnarray}
%
\begin{figure}[ht]
\vspace*{-0.2cm}
\begin{center}
\epsfig{file=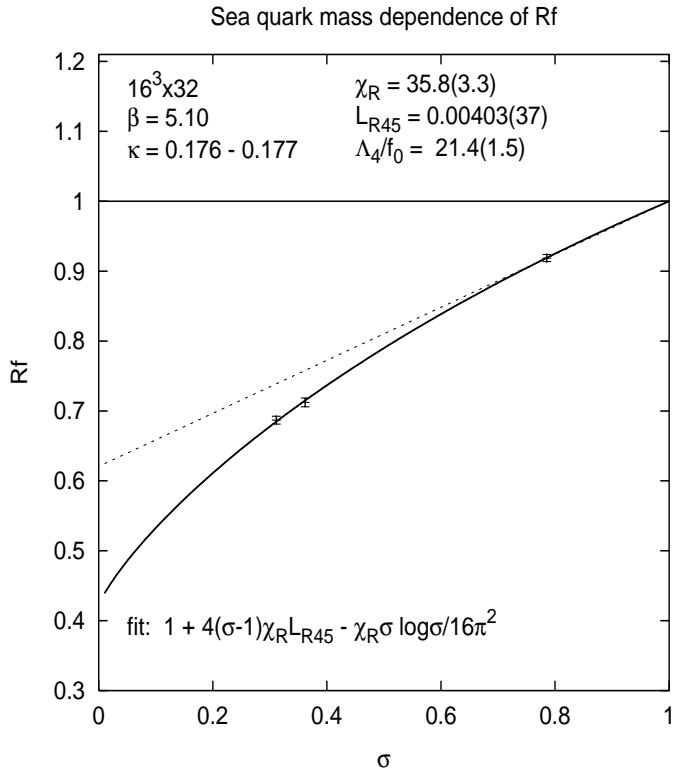,
        width=105mm,height=140mm,angle=-90,
        bbllx=50pt,bblly=-60pt,bburx=554pt,bbury=690pt}
\parbox{12cm}{\caption{
 Sea quark mass dependence of the pion decay constant.
 The straight dashed line connects the first two points.
\label{fig1}}}
\end{center}
\end{figure}

\begin{figure}[ht]
\vspace*{-0.2cm}
\begin{center}
\epsfig{file=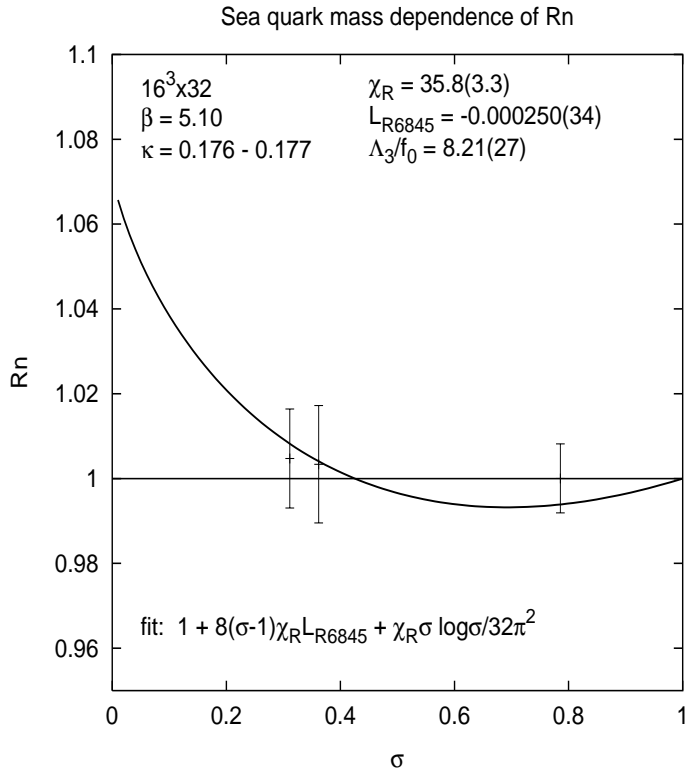,
        width=105mm,height=140mm,angle=-90,
        bbllx=50pt,bblly=-60pt,bburx=554pt,bbury=690pt}
\parbox{12cm}{\caption{
 Sea quark mass dependence of the pion mass-squared divided by the
 quark mass.
\label{fig2}}}
\end{center}
\end{figure}

 Consistent results with smaller errors can be obtained if one takes
 the value of $\chi_R=35.8(3.3)$ from the fit of the valence quark mass
 dependences (see next section) and performs two linear fits with
 the parameters $\chi_R L_{R45}$ and $\chi_R L_{R6845}$, respectively.
 The resulting parameters are
\begin{eqnarray}
&& \chi_R L_{R45} = 0.1443(15) \ ,      \hspace{1em}
L_{R45} = 4.03(37) \cdot 10^{-3} \ ,    \hspace{1em}
\frac{\Lambda_4}{f_0} = 21.4(1.5) \ ,
\nonumber
\\[0.5em]\label{eq2-9:09}
&& \chi_R L_{R6845} = -0.00896(86) \ ,  \hspace{1em}
L_{R6845} = -2.50(34) \cdot 10^{-4} \ , \hspace{1em}
\frac{\Lambda_3}{f_0} = 8.21(27)
\end{eqnarray}
 and the fits are shown in figures \ref{fig1} and \ref{fig2}.

 As these figures show, both $Rf_{SS}$ and $Rn_{SS}$ can be well
 fitted with the NLO ChPT formula.
 The fit parameters are within the expected range.
 For instance, the value of $\chi_R$ is rather close to the tree-level
 estimate $\chi_R^{estimate} \approx M_r/(r_0 f_0)^2 \simeq 40.3$.
 (Here we used $r_0f_0 \simeq 0.23$.)
 The presence of a {\em chiral logarithm} which causes the curvature
 is clearly displayed in figure \ref{fig1} where a straight line
 connecting the first two points is also shown.
 In $Rn_{SS}$ the measured points are consistent with the presence of
 a chiral logarithm but the relative errors are large because all the
 values including the ChPT fit are very close to 1.
 This implies that the deviation from the tree-level behaviour
 $m_{SS}^2 \propto \chi_S$ is rather small.
 The results for the parameters in (\ref{eq2-9:09}) are close to the
 ones reported in \cite{SEA}: the values for $\Lambda_4/f_0$
 practically coincide and the value of $\Lambda_3/f_0$ is only slightly
 higher now.

 The extrapolated values of $Rf_{SS}$ and $Rn_{SS}$ at zero quark mass
 are, respectively:
\be\label{eq2-10:10}
Rf_0 = 0.4228(60) \ ,  \hspace{3em}
Rn_0 = 1.0717(69) \ .
\ee
 The value of $Rf_0$ togehter with $Z_A^{-1} a f_\pi$ from table
 \ref{tab1} and $r_0=0.5\,{\rm fm}$ imply for the pion decay constant at
 zero quark mass ($f_0$)
\be\label{eq2-11:11}
Z_A^{-1} f_0 = 121(5)\,{\rm MeV} \ .
\ee
 This result for $N_s=2$ light quarks compares well with the
 phenomenological value $f_0=93\,{\rm MeV}$ if, as expected,
 $Z_A = {\cal O}(1)$.

\section{Valence quark mass dependence}\label{sec3}

 We consider for fixed sea quark mass $\chi_S$ the valence quark mass
 dependence of $f_\pi$ and $m_\pi^2$ as a function of the quark mass
 ratio
\be\label{eq3-1:12}
\xi \equiv \frac{\chi_V}{\chi_S} \ .
\ee
 In our numerical data we determined the pseudoscalar mass and decay
 constant in relatively wide ranges of the valence quark mass ratios,
 typically $\half \leq \xi \leq 2$.
 At the smaller quark masses ($\kappa=\kappa_{2,3}$), however, 
 for $\xi < 1$ {\em exceptional gauge configurations} appear which blow
 up the statistical errors and clearly influence the mean values
 themselves.
 Therefore, in most cases we restrict our fits to valence quark masses
 larger than the sea quark mass ($\xi > 1$).

 In the partially quenched situation several types of ratios can be
 constructed because the pseudoscalar meson can be the bound state of
 two valence quarks ($VV$) and also a valence quark and a sea quark
 ($VS$).
 The PQChPT formulas for the ratios of decay constants are:
\begin{eqnarray}
Rf_{VV} \equiv \frac{f_{VV}}{f_{SS}}
&=& 1 + 4(\xi-1)\chi_S L_{S5} -
\frac{N_s\chi_S}{64\pi^2}(1+\xi)\log\frac{1+\xi}{2}
\nonumber
\\[1.0em]\label{eq3-2:13}
&&+\; D_{fVV} \chi_S^2 (\xi-1) + Q_{fVV} \chi_S^2 (\xi-1)^2
+ {\cal O}(\chi_S^2\log\xi,\chi_S^3) 
\end{eqnarray}
 and
\begin{eqnarray}
Rf_{VS} \equiv \frac{f_{VS}}{f_{SS}}
&=& 1 + 2(\xi-1)\chi_S L_{S5}
\nonumber
\\[1.0em]
&&+\; \frac{\chi_S}{64N_s\pi^2}(\xi-1-\log\xi)
- \frac{N_s\chi_S}{128\pi^2}(1+\xi)\log\frac{1+\xi}{2}
\nonumber
\\[1.0em]\label{eq3-3:14}
&&+\; \half D_{fVV} \chi_S^2 (\xi-1) + Q_{fVS} \chi_S^2 (\xi-1)^2
+ {\cal O}(\chi_S^2\log\xi,\chi_S^3) \ .
\end{eqnarray}
 The analogous formulas for the valence quark mass dependence of the
 (squared) pseudoscalar meson masses are:
\begin{eqnarray}
Rn_{VV} \equiv \frac{m_{VV}^2}{\xi m_{SS}^2}
&=& 1 + 8(\xi-1)\chi_S(2L_{S8}-L_{S5})
+ \frac{\chi_S}{16N_s\pi^2}[\xi-1+(2\xi-1)\log\xi]
\nonumber
\\[1.0em]\label{eq3-4:15}
&&+\; D_{nVV} \chi_S^2 (\xi-1) + Q_{nVV} \chi_S^2 (\xi-1)^2
+ {\cal O}(\chi_S^2\log\xi,\chi_S^3)
\end{eqnarray}
 and
\begin{eqnarray}
Rn_{VS} \equiv \frac{2m_{VS}^2}{(\xi+1) m_{SS}^2}
&=& 1 + 4(\xi-1)\chi_S(2L_{S8}-L_{S5})
+ \frac{\chi_S}{16N_s\pi^2}\xi\log\xi
\nonumber
\\[1.0em]
&&+\; \half D_{nVV} \chi_S^2 (\xi-1) + Q_{nVS} \chi_S^2 (\xi-1)^2
\nonumber
\\[1.0em]\label{eq3-5:16}
&&+\; {\cal O}(\chi_S^2\log\xi,\chi_S^3) \ .
\end{eqnarray}
 In these formulas the Gasser-Leutwyler coefficients
 $L_{Sk}\; (k=4,5,\ldots)$ are defined at the scale $f_0\sqrt{\chi_S}$
 and, in addition to the NLO terms, also the tree-graph
 (i.e.~counterterm) contributions of the NNLO are included.
 Their general form is taken from ref.~\cite{SHARPE-WATER} and is
 discussed in more detail in section 2.1 of \cite{SEA}.
 The left-out terms of NNLO, which come from two-loop integrals,
 are generically denoted here by ${\cal O}(\chi_S^2\log\xi)$.

 In addition to the {\em single ratios} $Rf_{VV},\;Rf_{VS},\;Rn_{VV}$
 and $Rn_{VS}$ it is useful to consider the so called {\em double
 ratios} which do not depend on any of the NLO coefficients $L_{Sk}$.
 The PQChPT formulas for the double ratios are:
\begin{eqnarray}
RRf \equiv \frac{f_{VS}^2}{f_{VV}f_{SS}} 
&=& 1 + \frac{\chi_S}{32N_s\pi^2}(\xi-1-\log\xi)
\nonumber
\\[1.0em]\label{eq3-6:17}
&&+\; Q_{fd} \chi_S^2 (\xi-1)^2 + {\cal O}(\chi_S^2\log\xi,\chi_S^3)
\end{eqnarray}
 and
\begin{eqnarray}
RRn \equiv \frac{4\xi m_{VS}^4}{(\xi+1)^2m_{VV}^2 m_{SS}^2}
&=& 1 - \frac{\chi_S}{16N_s\pi^2}(\xi-1-\log\xi)
\nonumber
\\[1.0em]\label{eq3-7:18}
&&+\; Q_{nd} \chi_S^2 (\xi-1)^2 + {\cal O}(\chi_S^2\log\xi,\chi_S^3) \ .
\end{eqnarray}

 In the PQChPT formulas (\ref{eq3-2:13})-(\ref{eq3-7:18}) there are
 altogether 11 parameters.
 Three of them appear at NLO, namely with $N_s=2$
\be\label{eq3-8:19}
\chi_R \ ,  \hspace{2em}
\chi_R L_{R5} \ ,  \hspace{2em}
\chi_R L_{R85} \equiv \chi_R (2L_{R8}-L_{R5})
\ee
 and the rest in NNLO:
\be\label{eq3-9:20}
\chi_R^2 D_{fVV,nVV} \ ,  \hspace{2em}
\chi_R^2 Q_{fVV,fVS,fd,nVV,nVS,nd} \ .
\ee
 At the smallest quark mass fits with the NLO formulas are reasonable
 but for the larger quark masses the NNLO contributions are required
 unless the fits are restricted to a small range around $\xi=1$.

 An acceptable global fit with 11 parameters can be achieved if the
 valence quark mass dependence at all four sea quark masses is
 simultaneously considered.
 In this case one has to choose a {\em reference sea quark mass}
 $\chi_R$ and take into account the relation between the NLO parameters
\be\label{eq3-10:21}
L_{Sk} = L_{Rk} - c_k\log\frac{\chi_S}{\chi_R}
\ee
 where the relevant constants are:
\be\label{eq3-11:22}
c_5 = \frac{1}{128\pi^2} \ , \hspace{3em}
c_{85} \equiv 2c_8 - c_5 = -\frac{1}{128\pi^2} \ .
\ee

 Fitting all six valence quark mass dependences
 ($Rf_{VV}$, $Rf_{VS}$, $RRf$, $Rn_{VV}$, $Rn_{VS}$, $RRn$) there are
 reasonably good 11 parameter (linear) fits with
 $\chi^2 \simeq 200 \simeq$ no. of degrees of freedom.
 A typical set of the resulting fit parameters is shown in table
 \ref{tab2}.
\begin{table}[ht]
\vspace*{-0.5cm}
\begin{center}
\parbox{12cm}{\caption{\label{tab2}\em
 Values of best fit parameters for the valence quark mass dependence.
 Quantities directly used in the fitting procedure are in boldface.}}
\end{center}
\begin{center}
\begin{tabular}{|c|r||c|r|}\hline
\bteq{\chi_R}           & $35.8(3.3)$   &           &
\\\hline
\bteq{\chi_RL_{R5}}     & $0.1003(76)$  & $L_{R5}$  & $2.80(39)\cdot10^{-3}$   
\\
\bteq{\chi_RL_{R85}}    & $-0.0256(12)$ & $L_{R85}$ & $-0.714(65)\cdot10^{-3}$  
\\\hline
\bteq{\chi_R^2D_{fVV}}  & $-0.109(42)$  & $D_{fVV}$ & $-8.5(4.4)\cdot10^{-5}$  
\\
\bteq{\chi_R^2Q_{fVV}}  & $-0.014(29)$  & $Q_{fVV}$ & $-1.1(2.3)\cdot10^{-5}$  
\\
\bteq{\chi_R^2Q_{fVS}}  & $-0.0177(94)$ & $Q_{fVS}$ & $-1.39(81)\cdot10^{-5}$  
\\
\bteq{\chi_R^2Q_{fd}}   & $-0.0180(31)$ & $Q_{fd}$  & $-1.41(13)\cdot10^{-5}$ 
\\\hline
\bteq{\chi_R^2D_{nVV}}  & $-0.134(21)$  & $D_{nVV}$ & $-10.46(93)\cdot10^{-5}$  
\\
\bteq{\chi_R^2Q_{nVV}}  & $-0.087(13)$  & $Q_{nVV}$ & $-6.77(30)\cdot10^{-5}$  
\\
\bteq{\chi_R^2Q_{nVS}}  & $-0.0394(44)$ & $Q_{nVS}$ & $-3.07(24)\cdot10^{-5}$  
\\
\bteq{\chi_R^2Q_{nd}}   & $0.0077(48)$  & $Q_{nd}$  & $0.60(26)\cdot10^{-5}$   
\\\hline
\end{tabular}
\end{center}
\end{table}

 Comparing table \ref{tab2} to the corresponding one (table 4) in
 ref.~\cite{SEA} one can see that most values are within statistical
 errors the same.
 This is also true for the NLO parameters defined at the scale
 $4\pi f_0$, which are now
\begin{eqnarray}
\alpha_5 &\equiv& 128\pi^2 L_{R5}+\log\frac{\chi_R}{16\pi^2} = 2.06(42)\ ,
\nonumber
\\[0.5em]\label{eq3-12:23}
\alpha_{85} \equiv 2\alpha_8 - \alpha_5 
&\equiv& 128\pi^2 L_{R85}-\log\frac{\chi_R}{16\pi^2} = 0.583(45) \ .
\end{eqnarray}
 The value of $\alpha_5$ is practically the same as in table 5 of
 \cite{SEA} whereas $\alpha_{85}$ is slightly smaller now.

\begin{figure}[ht]
\vspace*{-4mm}
\begin{flushleft}
\epsfig{file=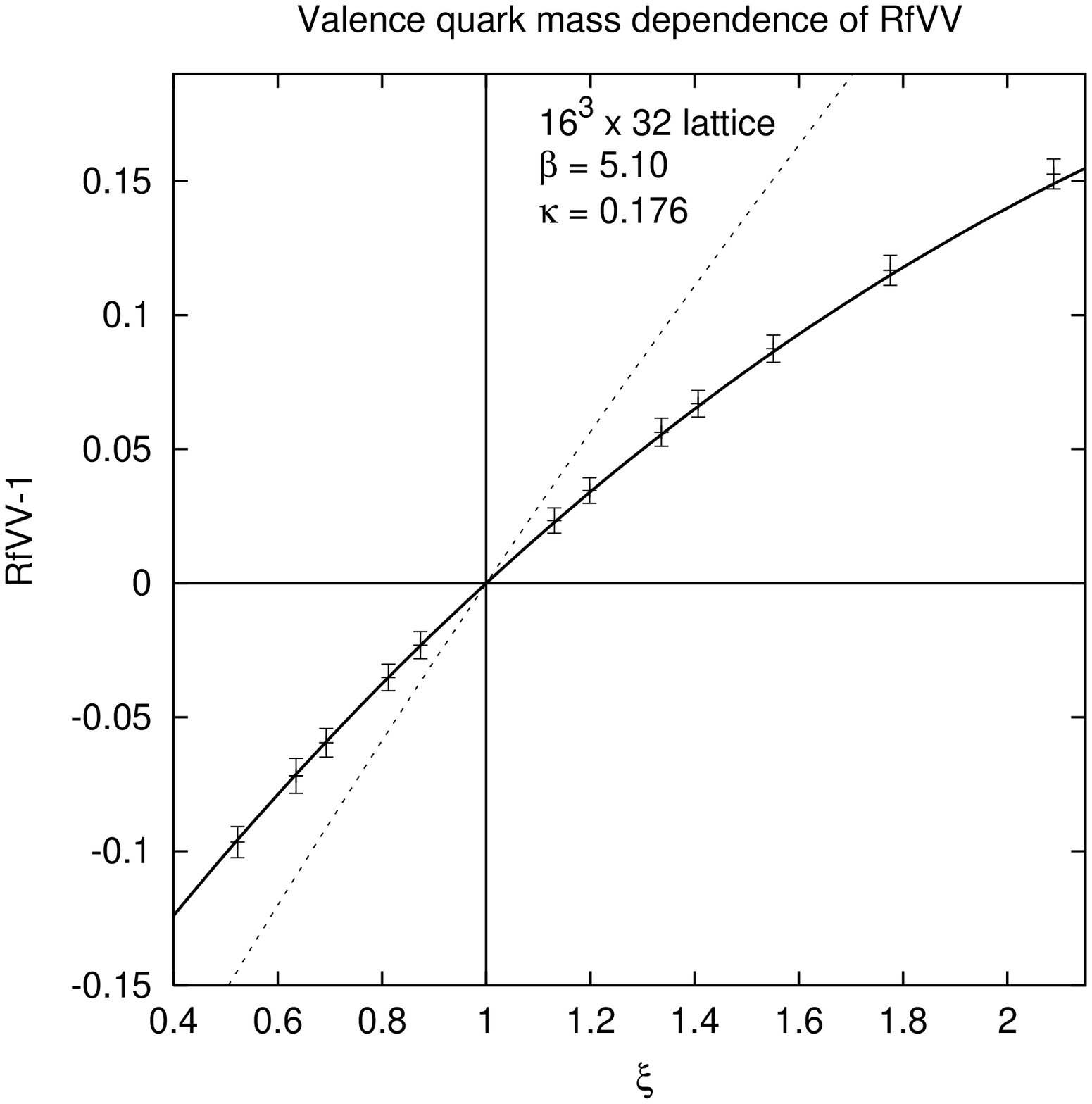,
        width=69mm,height=92mm,angle=-90,
        bbllx=50pt,bblly=-60pt,bburx=554pt,bbury=690pt}
\end{flushleft}
\vspace*{-78mm}
\begin{flushright}
\epsfig{file=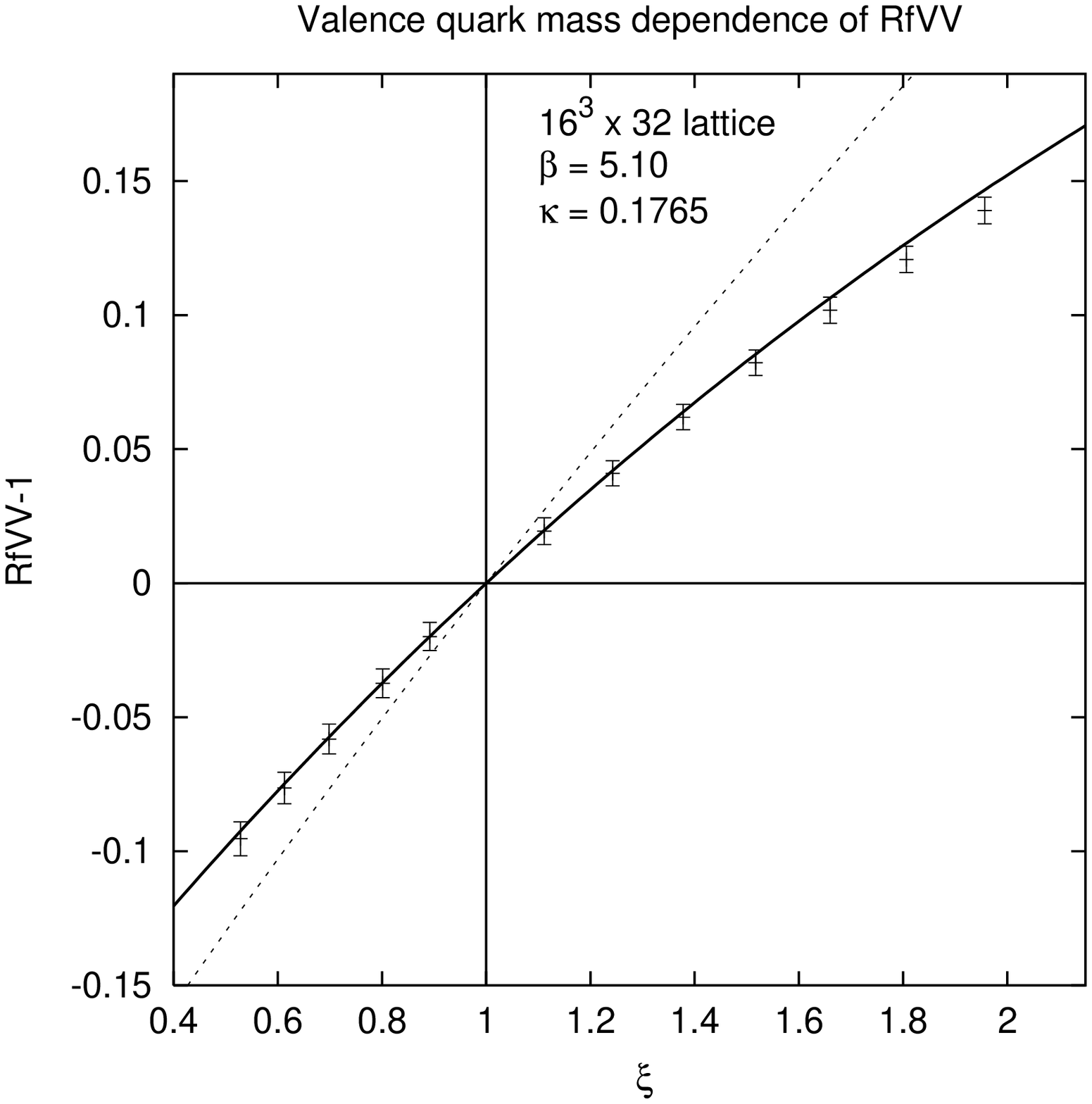,
        width=69mm,height=92mm,angle=-90,
        bbllx=50pt,bblly=-60pt,bburx=554pt,bbury=690pt}
\end{flushright}
\vspace*{-4mm}
\begin{flushleft}
\epsfig{file=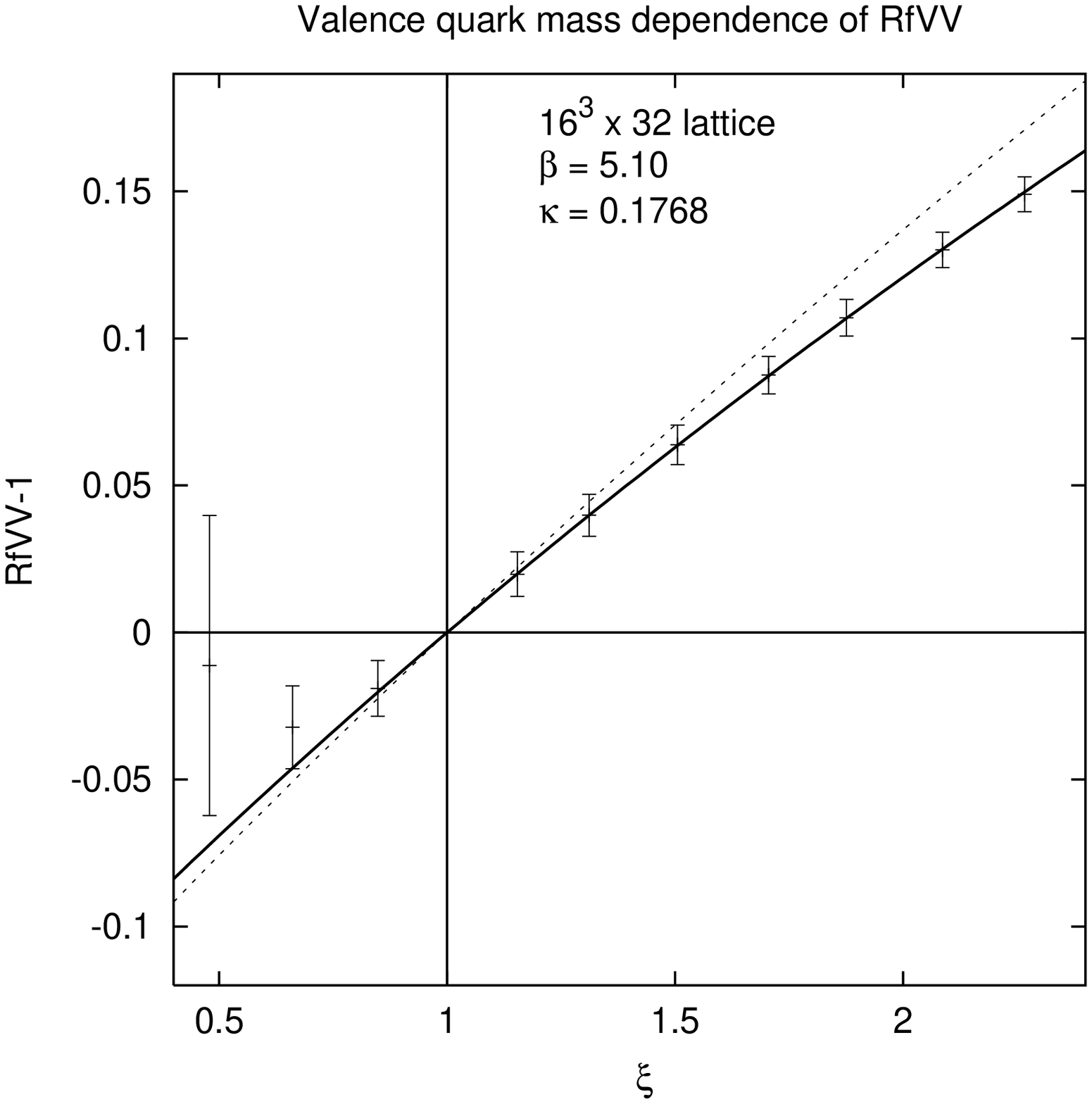,
        width=69mm,height=92mm,angle=-90,
        bbllx=50pt,bblly=-60pt,bburx=554pt,bbury=690pt}
\end{flushleft}
\vspace*{-78mm}
\begin{flushright}
\epsfig{file=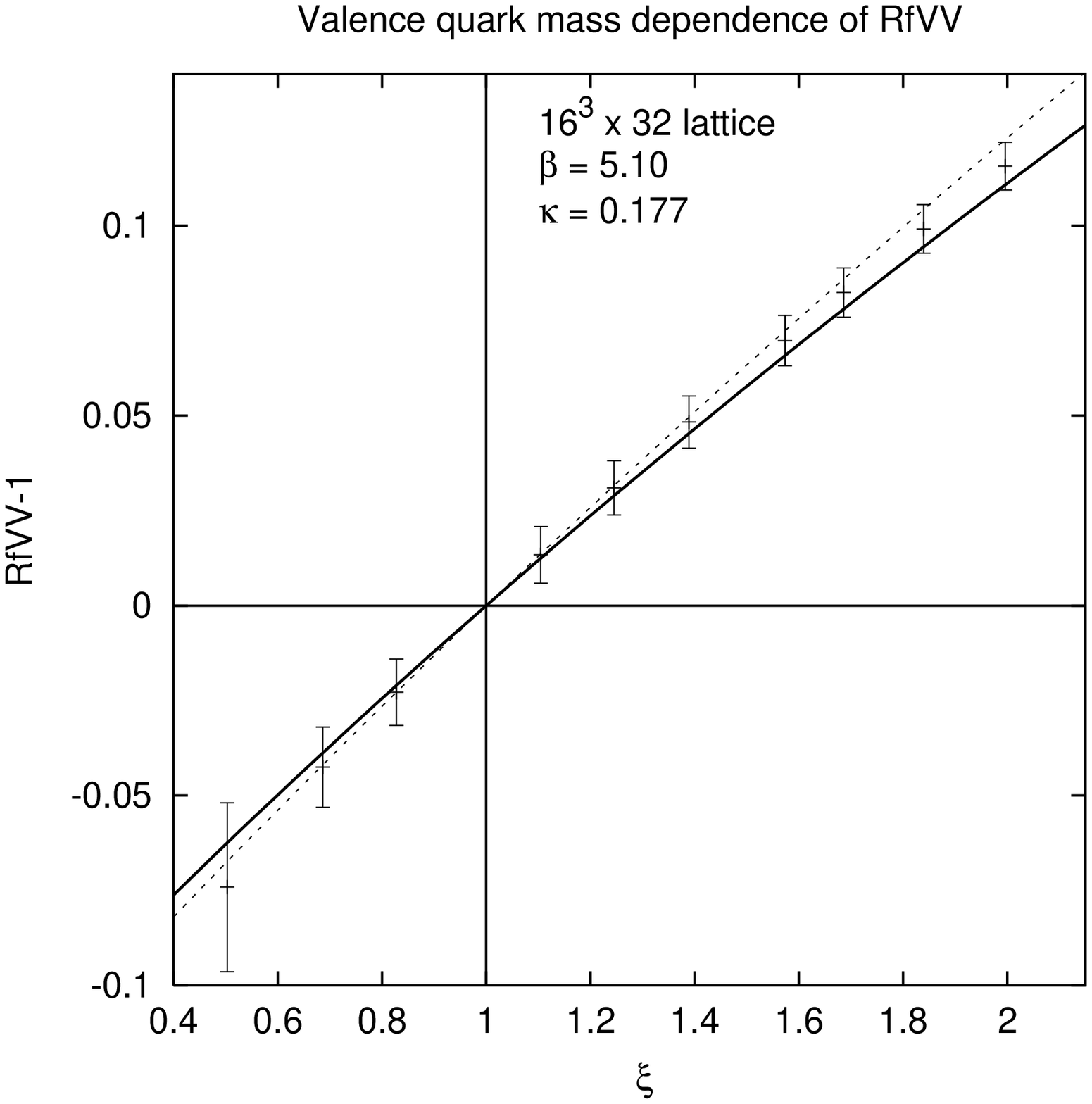,
        width=69mm,height=92mm,angle=-90,
        bbllx=50pt,bblly=-60pt,bburx=554pt,bbury=690pt}
\end{flushright}
\vspace*{-7mm}
\begin{center}
\parbox{12cm}{\caption{
 The linear fit for $Rf_{VV}$ for different sea quark masses.
 The NLO contributions alone are shown by dashed lines.
\label{fig4}}}
\end{center}
\vspace*{-5mm}
\end{figure}

 The tree-graph NNLO contributions play an important r\^ole in the
 global fits of the valence quark dependences, especially at the two
 larger sea quark mass ($\kappa=\kappa_0$ and $\kappa=\kappa_1$).
 At the two smaller sea quark masses NNLO is substantially less
 important.
 This is illustrated in figure \ref{fig3} where the 11 parameter fit
 for $Rf_{VV}$ is shown together with the NLO contributions alone.

\subsection{${\cal O}(a)$ terms}\label{sec3.1}

 The fits above have been performed with the continuum formulas --
 without ${\cal O}(a)$ or any other lattice artifacts.
 The fits are reasonably good and the resulting parameters are quite
 similar to those obtained in ref.~\cite{SEA} where the ${\cal O}(a)$
 terms have been taken into account in the (PQ)ChPT Lagrangian according
 to ref.~\cite{RUPAK-SHORESH}.
 It has been observed already in \cite{SEA} that the parameter in the
 chiral Lagrangian characterizing the magnitude of ${\cal O}(a)$
 effects
\be\label{eq3-13:24}
\rho \equiv \frac{2W_0 ac_{SW}}{f_0^2}
\ee
 is rather small compared to the quark mass parameter $\chi$ in
 (\ref{eq2-1:01}).
 Fitting the ratio $\eta_S \equiv \rho_S/\chi_S$ separately for the
 individual sea quark mass values we obtained increasing values for
 increasing sea quark masses: $0.02 \leq \eta_S \leq 0.07$.

\begin{figure}[ht]
\vspace*{-0.2cm}
\begin{center}
\epsfig{file=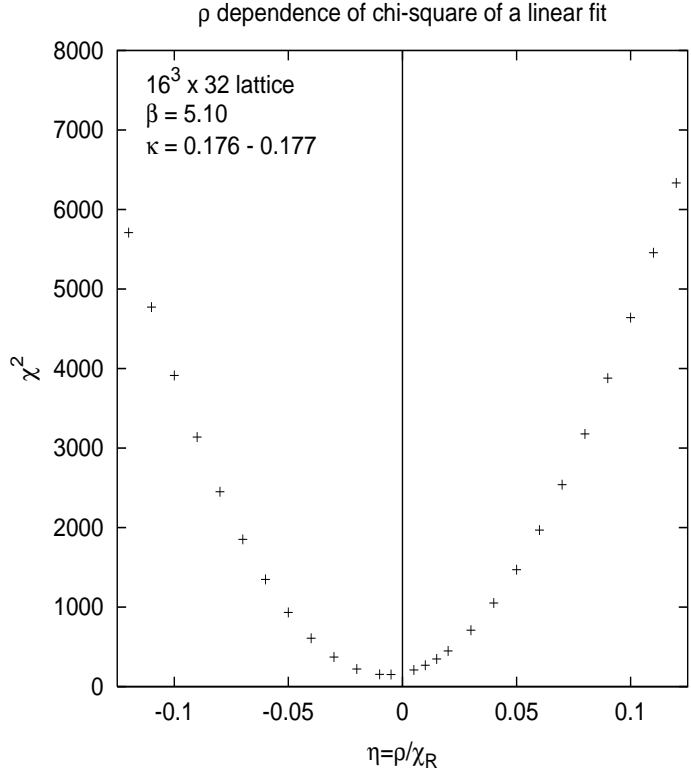,
        width=105mm,height=140mm,angle=-90,
        bbllx=50pt,bblly=-60pt,bburx=554pt,bbury=690pt}
\parbox{12cm}{\caption{
 $\chi^2$ of the linear fit of the ratios $Rf_{VV}$, $Rf_{VS}$, $RRf$,
 $Rn_{VV}$, $Rn_{VS}$ and $RRn$ for all four sea quark masses at fixed
 values of the ${\cal O}(a)$ parameter $\rho$ in the chiral Lagrangian.
\label{fig3}}}
\end{center}
\end{figure}

 The parameter $\rho$ should be independent of the quark mass because
 the quark masses are the other expansion parameters in the chiral
 Lagrangian.
 This means that a quark mass dependent $\rho_S$ incorporates some
 higher order effects proportional to some power of $am_q$.
 (For instance, a linearly incrasing value of $\eta_S$ corresponds to
 $\rho_S \propto (am_q)^2$.)
 Since the observed values of $\rho$ are small anyway it is interesting
 to consider the behaviour of the chi-square as a function of $\rho$ if
 the linear fits are performed for fixed $\rho$ .
 Because of the presence of another new parameter describing
 ${\cal O}(a)$ effects in the chiral Lagrangian, the linear fit has
 12 parameters for $\rho \ne 0$  (instead of 11 for $\rho=0$).
 As it is shown by figure \ref{fig3}, the $\chi^2$ of the fit has a
 minimum near $\rho=\eta=0$ and becomes extremely large already at
 $|\eta| \simeq 0.1$ where the absolute value of $\rho$ is 10\% of
 the value of the reference quark mass parameter $\chi_R$.

 Another way to investigate the importance of ${\cal O}(a)$ effects in
 our data is to consider the following combination of double ratios:
\be\label{eq3-14:25}
RRn+2\,RRf-3 = (Q_{nd}+2Q_{fd}) \chi_S^2 (\xi-1)^2 + 
{\cal O}(\chi_S^2\log\xi,\chi_S^3) \ .
\ee
 As this formula shows, this combination vanishes in next-to-leading
 order and only NNLO and higher orders contribute to it.
 On the lattice there could also be ${\cal O}(a)$ contributions which
 can be parametrized as
\begin{eqnarray}
RRn+2\,RRf-3 &=& 16\rho\,L_{S4W6} \frac{(\xi-1)^2}{\xi(\xi+1)}
- \rho\,\frac{(\xi-1)^2}{\chi_S\xi(\xi+1)}
\nonumber
\\[1.0em]
&&+\; \rho\,\frac{[2(1-\xi^2)+\log\xi+3\xi^2\log\xi]}{32\pi^2\xi(\xi+1)}
+ \rho\,\frac{(\xi-1-\xi\log\xi)}{32\pi^2\xi}
\nonumber
\\[1.0em]\label{eq3-15:26}
&&+\; {\cal O}(\rho^2,\chi^2) \ .
\end{eqnarray}
 Here only the linear piece of the $\eta_S=\rho/\chi_S$-dependence is
 kept because $\eta_S$ is small.
 $L_{S4W6} \equiv L_{S4}-W_{S6}$ is a new parameter appearing in the
 ${\cal O}(a)$ terms of the chiral Lagrangian \cite{RUPAK-SHORESH}.

\begin{figure}[ht]
\vspace*{-0.2cm}
\begin{center}
\epsfig{file=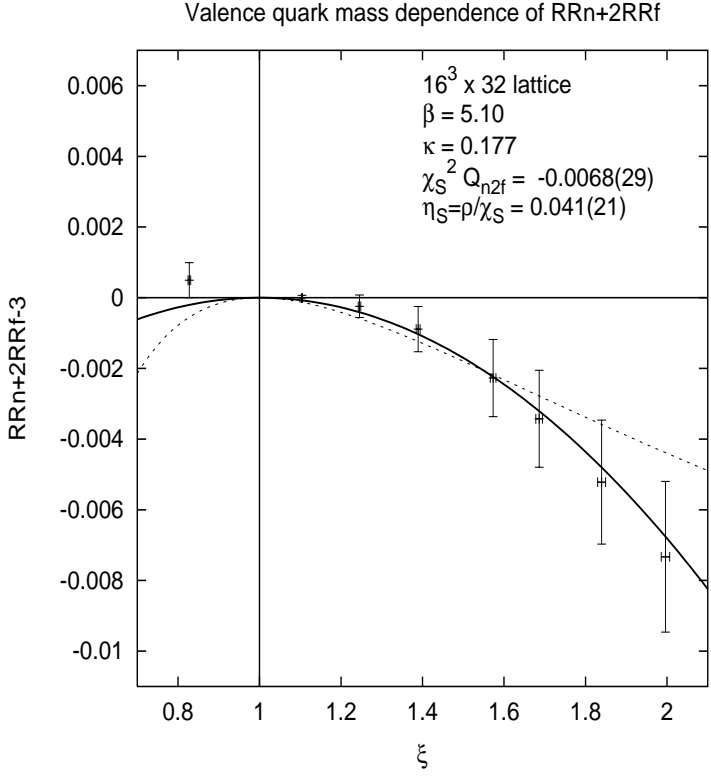,
        width=105mm,height=140mm,angle=-90,
        bbllx=50pt,bblly=-60pt,bburx=554pt,bbury=690pt}
\parbox{12cm}{\caption{
 Comparing the NNLO fit (full line) with the leading ${\cal O}(a)$ fit
 (dashed line) for $(RRn+2\,RRf-3)$ at $\kappa=\kappa_3$.
\label{fig5}}}
\end{center}
\end{figure}

 The linear fits with $\chi_S^2 Q_{n2f} \equiv \chi_S^2(Q_{nd}+2Q_{fd})$
 in (\ref{eq3-14:25}) and with $\rho$ in (\ref{eq3-15:26}), respectively,
 are shown in case of the smallest sea quark mass ($\kappa=\kappa_3$)
 by figure \ref{fig5}.
 As this figure shows, the NNLO fit with $\chi_S^2 Q_{n2f}$ is better
 ($\chi^2=1.3$) than the one with the leading ${\cal O}(a)$ term
 proportional to $\rho$ ($\chi^2=7.2$).
 For simplicity, the parameters $\chi_S=11.7$ and $L_{S4W6}=0.001$ are
 fixed in this latter case but taking other values does not change the
 qualitative picture.

 At the larger sea quark mass values the fits with the leading
 ${\cal O}(a)$ terms behave similarly to figure \ref{fig5}.
 This supports the fact that the ${\cal O}(a)$ terms are not important
 in our numerical data.
 As shown by figure \ref{fig3}, good fits can only be obtained at rather
 small values of $\eta=\rho/\chi$.
 In contrast, the NNLO contributions are very important especially at our
 larger sea quark masses.

\section{Studies of the updating algorithm}\label{sec4}

 The numerical simulations have been performed by the two-step
 multi-boson (TSMB) algorithm \cite{TSMB}.
 This dynamical fermion update algorithm is based on the
 {\em multi-boson representation} of the fermion determinant
 \cite{LUSCHER} and in its present form it incorporates several modern
 ideas of fermionic updating: the {\em global correction step} in the
 update \cite{BORFOR}, the final {\em reweighting correction}
 \cite{FREJAN} and the {\em determinant breakup} \cite{HASENAL}.

 Our error analysis is based on measuring the autocorrelations of the
 quatities in question \cite{ALPHA:BENCHMARK,WOLFF} therefore we can
 estimate the {\em computation cost} based on the integrated
 autocorrelations $\tau_{int}$.
 In our previous papers \cite{NF2TEST,PRICE,TSUKUBA} we proposed an
 approximate formula for the cost
\be\label{eq4-1:27}
C_{\tau_{int}} \simeq F\, (am_q)^{-2}\, \Omega 
\ee
 where $am_q$ is the quark mass in lattice units and $\Omega$ the number
 of lattice points.
 The overall factor $F$ depends on the quantity under investigation.
 If we count the cost in terms of the number of floating point
 operations necessary to perform an update sequence with length
 $\tau_{int}$ then the present simulations on $16^3 \cdot 32$ lattice
 are consistent with
\be\label{eq4-2:28}
F_{plaquette} \simeq 7\cdot 10^6 \ , \hspace{2em}
F_{m_\pi} \simeq 10^6 \ , \hspace{2em}
F_{f_\pi} < 4\cdot 10^5 \ .
\ee
 In case of $f_\pi$ we only have an upper limit on $\tau_{int}$ because
 the gauge configurations stored for the measurements were statistically
 independent.
 These numbers are somewhat smaller than our previous estimates in
 \cite{NF2TEST,PRICE,TSUKUBA} which is due to a better tuning of
 algorithmic parameters.
 In particular, these simulations were done with a determinant breakup
 $N_b=4$ which means that the fermion determinant of the two degenerate
 flavours ($N_f=2$) were reproduced by $4\otimes(N_f=\half)$ flavours.
 Another important point is the frequent call of the global heatbath
 update of the multi-boson fields which every time gives a
 statistically independent boson configuration.

 If we take the plaquette expectation value as the worst case, then
 at the present quark masses and lattice spacing this cost estimate is
 similar to previous estimates (see, for instance, the formula of
 A. Ukawa \cite{UKAWA}) but considering the more interesting cases of
 $m_\pi$ or $f_\pi$ there is a substantial improvement by an order of
 magnitude or more.
 In addition, towards large volumes, smaller quark masses and/or smaller
 lattice spacings the scaling of the cost estimate in (\ref{eq4-1:27})
 is better: for fixed lattice spacing the cost increases as
 $m_q^{-2}\Omega$ and decreasing the lattice spacing and keeping the
 physical parameters fixed the cost behaves as $a^{-6}$.
 This has to be compared to the estimated behaviour in \cite{UKAWA}
 $m_q^{-3}\Omega^{5/4}$ and $a^{-7}$, respectively.

\section{Summary}\label{sec5}

 The quark mass dependence of the pseudoscalar mass and decay constant
 in our numerical data can be well fitted with the continuum (PQ)ChPT
 formulas.
 It has been already observed on the $16^4$ lattice in ref.~\cite{SEA}
 that the ${\cal O}(a)$ lattice artifacts at our gauge coupling
 $\beta=5.1$, corresponding to a lattice spacing
 $a \simeq 0.2\,{\rm fm}$, are small and one can obtain reasonable fits
 by omitting them.
 This conclusion is strengthened by the new $16^3 \cdot 32$ data and
 therefore here we based our estimates of the chiral Lagrangian
 parameters on fits with the continuum formulas.

 The use of the ratios of the PCAC quark mass as the variable in
 comparing the simulation data to chiral perturbation theory is
 essential.
 Taking other quark mass definitions, for instance
 $\mu_{ren} \equiv Z_m(\mu_0-\mu_{cr})$, would be the source of large
 lattice artifacts at our lattice spacing.

 The sea quark mass dependence of $f_\pi$ and $m_\pi^2/m_q$ can be
 well described in our quark mass range $0.855 \leq M_r \leq 2.13$,
 which roughly corresponds to
 $\frac{1}{4}m_s \leq m_q \leq \frac{2}{3}m_s$, by the NLO ChPT
 formulas.
 The obtained estimates of the relevant Gasser-Leutwyler constants are,
 according to section \ref{sec2}:
\be\label{eq5-1:29}
\frac{\Lambda_3}{f_0} = 8.21(27) \ , \hspace{4em}
\frac{\Lambda_4}{f_0} = 21.4(1.5) \ .
\ee
 The functional dependence of the ratio of $f_\pi$ as a function of
 the ratio of quark masses clearly shows the presence of chiral
 logarithms (see figure \ref{fig1}).
 This observation is in agreement with the results in a recent paper of
 the UKQCD Collaboration \cite{UKQCD:LOG} which came out during writing
 up this paper.

 In the valence quark mass dependence of the same quantities, in
 addition to the NLO terms, the higher order NNLO contributions appear
 to be important -- especially at our two larger sea quark masses.
 But, as shown by figure \ref{fig3}, the importance of the NNLO terms is
 considerably reduced at the two lighter sea quark masses.
 Our best estimates for the relevant Gasser-Leutwyler constants at the
 scale $4\pi f_0$ are, according to (\ref{eq3-12:23}):
\be\label{eq5-2:30}
\alpha_5 = 2.06(42) \ , \hspace{3em}
2\alpha_8 - \alpha_5 = 0.583(45) \ .
\ee

 The errors quoted in (\ref{eq5-1:29}) and (\ref{eq5-2:30}) are only the
 statistical ones.
 In order to decrease the systematic errors simulations at still smaller
 sea quark masses would be useful.
 Since our lattice volume is relatively large ($L \simeq 3\,{\rm fm}$),
 finite volume effects can be expected to be small (see
 \cite{COLDUR,BECVIL}).
 For the moment we have no direct handle on the magnitude of the
 remaining non-zero lattice spacing effects.
 These should be determined by performing simulations at smaller
 lattice spacings.

\vspace*{1em}\noindent
{\large\bf Acknowledgments}

\noindent
 The computations were performed on the APEmille systems installed 
 at NIC Zeu\-then, the Cray T3E systems at NIC J\"ulich, the PC
 clusters at DESY Hamburg and at the University of M\"unster and the Sun
 Fire SMP-Cluster at the Rechenzentrum - RWTH Aachen.

 We thank Stephan D\"urr, Karl Jansen and Gernot M\"unster for helpful
 suggestions and discussions.
 We thankfully acknowledge the contributions of Luigi Scorzato in the
 early stages of this work.

\newpage


\end{document}